\documentclass[review,10pt]{elsarticle}

\usepackage[a4paper,left=3.0cm,right=2.0cm,top=3.0cm]{geometry}

\usepackage[utf8]{inputenc}
\usepackage{listings}
\usepackage{xcolor}
\usepackage{hyperref}
\usepackage{svg}

\hypersetup{colorlinks=true, linkcolor=blue, filecolor=blue, urlcolor=blue, citecolor=blue}

\definecolor{codegreen}{rgb}{0,0.6,0}
\definecolor{codegray}{rgb}{0.5,0.5,0.5}
\definecolor{codepurple}{rgb}{0.58,0,0.82}
\definecolor{backcolour}{rgb}{0.95,0.95,0.92}

\lstdefinestyle{mystyle}{
  backgroundcolor=\color{backcolour},   commentstyle=\color{codegreen},
  keywordstyle=\color{magenta},
  numberstyle=\tiny\color{codegray},
  stringstyle=\color{codepurple},
  basicstyle=\ttfamily\scriptsize,
  breakatwhitespace=false,         
  breaklines=true,                 
  captionpos=b,                    
  keepspaces=true,                 
  numbers=left,                    
  numbersep=3pt,                  
  showspaces=false,                
  showstringspaces=false,
  showtabs=false,                  
  tabsize=1
}

\usepackage{graphicx}
\usepackage{amssymb}

\usepackage{lineno}

\journal{Computers \& Geosciences}

\begin{document}

\begin{frontmatter}


\title{A workflow for seismic imaging with quantified uncertainty}



\author[1,2]{Carlos H. S. Barbosa\fnref{label1}}
\author[4]{Liliane N. O. Kunstmann\fnref{label2}}
\author[1,2]{R\^omulo M. Silva\fnref{label3}}
\author[1]{Charlan D. S. Alves\fnref{label4}}
\author[2]{Bruno S. Silva\fnref{label5}}
\author[6]{Djalma M. S. Filho\fnref{label6}}
\author[4]{Marta Mattoso\fnref{label7}}
\author[3]{Fernando A. Rochinha\fnref{label1}}

\author[1]{Alvaro L.G.A. Coutinho\corref{cor1}\fnref{label7}}
\cortext[cor1]{Corresponding author:}
\ead{alvaro@nacad.ufrj.br}
\ead[orcid]{https://orcid.org/0000-0002-4764-1142}

\address[1]{High-Performance Computing Center, COPPE, Federal University of Rio de Janeiro, Brazil}
\address[2]{Department of Civil Engineering, COPPE, Federal University of Rio de Janeiro, Brazil}
\address[3]{Department of Mechanical Engineering, COPPE, Federal University of Rio de Janeiro, Brazil}
\address[4]{Department of Computer Science, COPPE, Federal University of Rio de Janeiro, Brazil}
\address[6]{CENPES, Petrobras}

\fntext[label1]{ Drafting the manuscript, coding the scientific workflow, and analysis of the seismic images and uncertainty maps.}
\fntext[label2]{ Drafting the manuscript, provenance coding, and analysis of the provenance results.}
\fntext[label3]{ HPC optimization and performance analysis.}
\fntext[label4]{ Coding the Bayesian Eikonal tomography and analysis of the results.}
\fntext[label5]{ Provided the necessary data, and seismic data analysis.}
\fntext[label6]{ Revising the manuscript critically for important intellectual content.}
\fntext[label7]{ Analysis and interpretation of data, drafting the manuscript, and revising the manuscript critically for important intellectual content.}

\begin{abstract}
The interpretation of seismic images faces challenges due to the presence of several uncertainty sources. Uncertainties exist in data measurements, source positioning, and subsurface geophysical properties.  Understanding uncertainties' role and how they influence the outcome is an essential part of the decision-making process in the oil and gas industry. Geophysical imaging is time-consuming. When we add uncertainty quantification, it becomes both time and data-intensive. In this work, we propose a workflow for seismic imaging with quantified uncertainty. We build the workflow upon Bayesian tomography, reverse time migration, and image interpretation based on statistical information. The workflow explores an efficient hybrid parallel computational strategy to decrease the reverse time migration execution time. High levels of data compression are applied to reduce data transfer among workflow activities and data storage. We capture and analyze provenance data at runtime to improve workflow execution, monitoring, and debugging with negligible overhead. Numerical experiments on the Marmousi2 Velocity Model Benchmark demonstrate the workflow capabilities. We observe excellent weak and strong scalability, and results suggest that the use of lossy data compression does not hamper the seismic imaging uncertainty quantification.
\end{abstract}


\end{frontmatter}


\vspace{1.0mm}

\section{Introduction}
\label{S:1}
An application of seismic imaging is delineating the structural geologic aspects of the subsurface. A migrated image is inferred from seismic data after careful processing and migration that intends to collapse diffraction and relocates seismic reflections events to their correct position in depth. The generation of images for seismic interpretation needs a migration tool, which has as inputs seismograms and an estimation of the spatial velocity distribution. This data is acquired from seismic measurements with appropriated pre-processing and velocity analysis techniques, like Normal Move-Out, tomography, or full-waveform inversion (FWI). 

In seismic exploration, many decisions rely on interpretations of seismic images, which are affected by multiple sources of uncertainty. The leading causes of uncertainty in seismic imaging are the geometry survey, source signature, data noise, data processing, parameters, and model and numerical errors \citep{caers2011, FomelLanda2014}. Such procedures lead, unavoidably, to uncertainties on the data that propagate to the velocity analysis and migration process. The effects of such uncertainties are hard to quantify. Thus, fundamental to the decision-making process is understanding uncertainties and how they influence the outcomes. Building a depth seismic migrated image is challenging due to the non-uniqueness of the inverse problem to estimate the uncertain velocity parameters.

Several works discuss how to express the uncertainty related to velocity-model building and measure the impact on the migration techniques \citep{caers2011, FomelLanda2014, Lietal2015, PoliannikocMalcolm2016, Messudetal2017, Elyetal2018}. Typically, uncertainties are characterized through a probabilistic perspective and expressed as an ensemble of possible realizations of a random variable, making the Monte Carlo (MC) method the standard tool to manage and quantify uncertainties \citep{BodinSambridge2009, Botteroetal2016, Belhadjetal2018, Michelioudakisetal2016, martin2012stochastic, rawlinson2014seismic}. Furthermore, these works emphasize the computational difficulties in quantifying the uncertainties, particularly those associated with the high dimensionality of the data. For instance, \cite{Lietal2015} and \citet{PoliannikocMalcolm2016} choose smart sampling strategies to overcome the computational limits of calculating posteriors for more realistic problems. We recall that to access uncertainties is necessary to migrate the recorded data in several hundreds of samples. Migrating the samples makes the computational algorithms not only time but also data intensive. Even in today's supercomputers, their implementation and data management become critical for the effective implementation of seismic imaging with quantified uncertainty.

To tackle the computational and data management challenges on seismic imaging with quantified uncertainty, we introduce a workflow built upon three sequential stages: (i)  Bayesian tomography to estimate the seismic velocity from first arrival travel times, where the forward modeling relies on the Eikonal equation (called here Bayesian Eikonal Tomography - BET), and the Reversible Jump Monte Carlo Markov Chain (RJMCMC) algorithm \citep{Green1995, GreenHastie2009, Belhadjetal2018}; (ii) Seismic migration using Reverse Time Migration (RTM) wrapped in a Monte Carlo algorithm; (iii) Image interpretation supported by statistical information. The BET inversion for velocity estimation offers the flexibility of treating any prior information, like a reasonable initial velocity estimation, and it has the advantage of providing results that help to quantify the uncertainties associated with the solutions \citep{Belhadjetal2018}. RTM migrates the seismograms for the velocity models from the first stage (BET) to generate the corresponding set of migrated seismic images. Lastly, we expose the uncertainty maps that give a global representation of the ensemble of velocity fields, and images produced respectively by BET and RTM.

Model-selection strategies \citep{Lietal2015} can decrease the number of models for the migration stage. Nevertheless, even in this case, RTM migration wrapped into the Monte Carlo method is still challenging due to the high dimensional uncertain inputs, the amount of data to manage, and computational costs associated with imposing the Courant-Friedrichs-Lewy (CFL) condition for the two-way wave equation. Therefore, we present a parallel solution for RTM with quantified uncertainty to mitigate the computational costs. Our implementation explores MPI to manage each RTM among distributed platforms, and OpenMP+vectorization to explore parallelism in hybrid architectures. Besides, data compression \citep{Lindstrometal2016} improves data transfer through the workflow stages, to read the seismogram from disk, to manage temporary files, and to store the results in the final stage. Data compression is necessary because the workflow generates a large amount of heterogeneous data. Finally, to manage and track data exchange among the stages, we capture provenance data to monitor the several file transformations in addition to registering CPU, file paths, and relationships among files.

This work is organized by initially briefly reviewing the essential components to build the workflow. After that, we detail the workflow with an emphasis on providing uncertainty estimations on the results. We discuss in the sequence high-performance computing improvements for RTM, and the additional data management tools for data compression and provenance. The paper ends with a summary of our main conclusions.

\section{Migration with Quantified Uncertainty}
 \label{sec:theory}
In this section, we present the main concepts supporting our probabilistic framework for seismic imaging with quantified uncertainty. Bayesian inference states the solution of an inverse problem as {\it a posteriori} probability distribution (pdf) \citep{Smith1991} of a random variable (often random fields), which in our case is a function that describes the spatial distribution of the heterogeneous subsurface velocity given the observed data. However, the posterior distribution cannot be expressed in a convenient analytical form \citep{BodinSambridge2009}. Hence, to address this problem, we use a Markov Chain Monte Carlo (MCMC) method to generate samples of the stochastic field. The samples ensemble serves as input data for generating seismic images, producing a migration with quantified uncertainty.

\subsection{Bayesian Eikonal Tomography}\label{sec:tomography}

Seismic tomography is an inversion tool that aims to identify physical parameters (velocity, anisotropy, among others) given a set of observations (measurement data).  Parameters and the data are related by the following equation:
\begin{eqnarray} \label{eq.seistomo}
\textbf{d}^{obs} = \textbf{F}(\textbf{m}) + \epsilon,
\end{eqnarray}
where $\epsilon$ is an additive noise including modeling and measurements errors. Here, the operator $\textbf{F}$ is the Eikonal equation, that connects the velocity with first arrival travel times \citep{Brantut2018}. The high-dimensional parameter vector $\textbf{m} = (\textbf{V}, \textbf{U}, n)$, where $n$ is the number of grid points, the $n$-dimensional vector $\textbf{U} = \{u_i\}$, contains the grid positions, and the vector $\textbf{V} = \{v_i\}$, $i=1,2, \ldots n$, the velocity of the compressional wave in the $i$-th computational grid point.

Iterative linearized approaches for seismic tomography have problems related to non-linearity and ill - posedness. These methodologies usually produce inaccurate parameters due to local minimum convergence. Besides, iterative linearized methods are not able to properly quantify the uncertainty associated with their solution. To address this issue, we have chosen a Bayesian formulation to aggregate as information as possible to yield a better parameter estimation and the associated uncertainties. A Bayesian approach aims at estimating the parameters conditional distribution with respect to known  data, through a combination of prior knowledge regarding the velocity model and information provided by such data. Thus, to define the posterior distribution $P(\textbf{m} | \textbf{d}^{obs})$, we use Bayes' theorem:
\begin{eqnarray} \label{eq.bayes}
P(\textbf{m} | \textbf{d}^{obs}) = \frac{P(\textbf{d}^{obs} | \textbf{m}) P(\textbf{m})}{P(\textbf{d}^{obs})},
\end{eqnarray}
where $P(\textbf{d}^{obs} | \textbf{m})$ is the likelihood function, $P(\textbf{m})$ is the {\it a priori} parameter density distribution, and $P(\textbf{d}^{obs})$ is the evidence.

The likelihood represents how well the estimated travel times fits the measurements. Assuming that the additive noise in Equation \ref{eq.seistomo} is Gaussian with zero mean and variance $\sigma_{d}^{2}$ the likelihood is:
\begin{eqnarray} \label{eq.likelihood}
P(\textbf{d}^{obs} | \textbf{m}) = exp \left( - \frac{1}{2} \left\| \frac{F(\textbf{m}) - \textbf{d}^{obs}}{\sigma_{d}^{2}} \right\|^{2} \right),
\end{eqnarray}
where $\| \cdot \|$ is the standard $l_{2}$-norm.

To explore the posterior distribution and deal with the space parameter dimension, we employ the Reversible Jump Markov Chain Monte Carlo (RJMCMC) method \citep{Green1995, GreenHastie2009, BodinSambridge2009, Belhadjetal2018}. The MCMC method is an iterative stochastic approach to generate samples according to the posterior probability distribution.  A generalized version, called Reversible Jump, allows inference on both parameters and its dimensionality \citep{BodinSambridge2009}. Thus, following \cite{Belhadjetal2018}, we employ a Voronoi's tesselation with Gaussian kernels as basis functions to constrain the inverse problem solution and reduce the parameter dimensionality. Such parametrization works as a prior that embodies the expected velocity field non-stationary character, displaying localized abrupt changes with a continuous variation. Hence, the number of Voronoi cells $n' \ll n$. Therefore, given the current state of the chain $\textbf{m}^{i}$, three perturbations are possible, that is, the probability of adding ($p_{a}$) new cell nuclei, probability of deleting ($p_{d}$) one existing cell nuclei, and the probability of moving ($p_{m}$) selected cell nuclei. Hence, the acceptance probability of the three perturbations depends on the ratio of their likelihoods. After a burn-in phase, the process iterates, eventually reaching the total number of pre-defined iterations. By attributing a constant velocity magnitude inside each Voronoi cell, we put a probabilistic structure in the prior, assuming that such velocity values follow independent uniform variables. Playing with the bound of the random variables, we can incorporate in the priors spatial trends learned by experts by different data sources, like the deposition history.

\subsection{Reverse Time Migration}\label{sec:rtm}

Reverse Time Migration (RTM) is a depth migration approach based on the two-way wave equation and an appropriate imaging condition. Restricting ourselves to the acoustic isotropic case, the wave equation is,
\begin{eqnarray} \label{eq.secondOrder}
\nabla^{2} p(\textbf{r},t) - \frac{1}{v^{2}(\textbf{r})}\frac{\partial^{2} p(\textbf{r},t)}{\partial t^{2}} = 0,
\end{eqnarray}
where $\textbf{r} = (r_{x}, r_{y}, r_{z})$ is the position vector, $v$ is the compressional wave velocity, and $p(\textbf{r},t)$ the hydrostatic pressure in the position $\textbf{r}$ and time $t$.

The RTM imaging condition \citep{Zhouetal2018} is the zero-lag cross-correlation of the forward-propagated source wavefield ($S$) and the backward-propagated receiver wavefield ($R$) normalized by the square of the source wavefield, 
\begin{eqnarray} \label{eq.imageCondition}
I(\textbf{r}) = \frac{\int_{T} S(\textbf{r},t) R(\textbf{r},t) dt}{\int_{T} S^{2}(\textbf{r},t) dt}.
\end{eqnarray}

The source wavefield is the solution of equation \ref{eq.secondOrder} with the seismic wavelet treated as a boundary condition. Similarly, the receiver wavefield is the solution of the same equation \ref{eq.secondOrder} with the observed data (seismograms) as a boundary condition. The normalization term in equation \ref{eq.secondOrder} produces a source-normalized cross-correlation image that has the same unit, scaling, and sign as the reflection coefficient \citep{Zhouetal2018}. Therefore, within the MC method, RTM migrates the near-surface recorded signals for the ensemble of BET-generated velocity samples.

\subsection{Uncertainty in Images} 
\label{sec:uncertaintymaps}

To reach a global representation of the solution variability within the ensemble of plausible velocity fields and images produced by the workflow, we follow \cite{LiSun2016}. They proposed a confidence index, that normalizes the uncertainty map explored by \citet{Lietal2015}, that assigns at each spatial point an evaluation of the degree of uncertainty expressed by the standard deviation $\sigma(\textbf{r})$, where low values represent regions where there are high variabilities, and high values the regions where there are small variabilities. The confidence index is expressed by,
\begin{eqnarray} \label{eq.confidence}
c(\textbf{r}) = \frac{\sigma_{max} - \sigma(\textbf{r})}{\sigma_{max} - \sigma_{min}},
\end{eqnarray}
where $c(\textbf{r})$ is the confidence in position $\textbf{r}$, and $\sigma_{min}$, $\sigma_{max}$ and are the minimum, maximum and standard deviations in the $\textbf{r}$ position, respectively.

An alternative measure is the ratio between standard deviation and average, that is,
\begin{eqnarray} \label{eq.ratioStdAvg}
r(\textbf{r}) = \frac{\sigma(\textbf{r})}{\mu(\textbf{r})},
\end{eqnarray}
where, $\mu$ represents the average map of the data. In this case, Equation \ref{eq.ratioStdAvg} measures the variations among the samples with the mean as reference.

\section{A Workflow for Seismic Imaging}
\label{sec:methodology}
In this section, we present the components of our workflow for seismic imaging with quantified uncertainty. We have chosen a particular set of chained components that provide a comprehensive probabilistic framework for seismic imaging. Nevertheless, we recognize that different components could be chained in a similar form to provide other possible workflows. Thus, a workflow is an abstraction that represents stages of computations as a set of activities and a data flow among them. Each activity can be a software that performs operations on an input data set, transforming it into an output data set. A workflow can represent the whole seismic imaging stages as a chain of activities. Stages of seismic imaging (or workflow activities) can be replaced or extended to build a different viewpoint of the outcomes, such as statistic information for uncertainty quantification.

\begin{figure}[ht]
\centering\includegraphics[scale=.3]{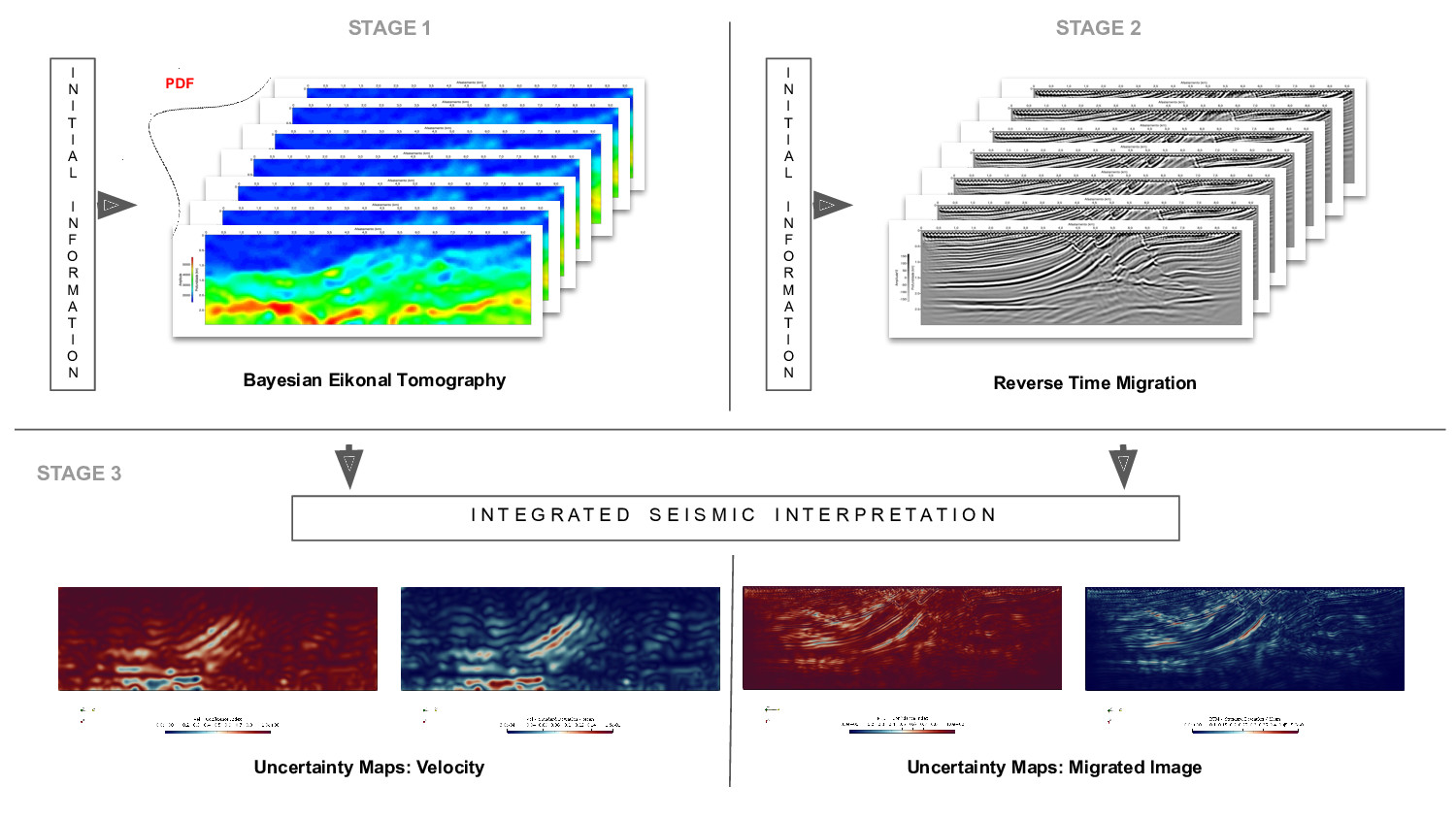}
\caption{Results of BET (Stage 1), and RTM for each velocity field (Stage 2) built-in Stage 1. Stage 3 shows the uncertainty maps for velocities and migrated images. The maps can guide interpreters to potential areas with low/high uncertainty in velocities and structures.}
\label{FIG:WOKFLOWRTMIMAGES}
\end{figure}

\begin{figure}[ht]
\centering\includegraphics[scale=.3]{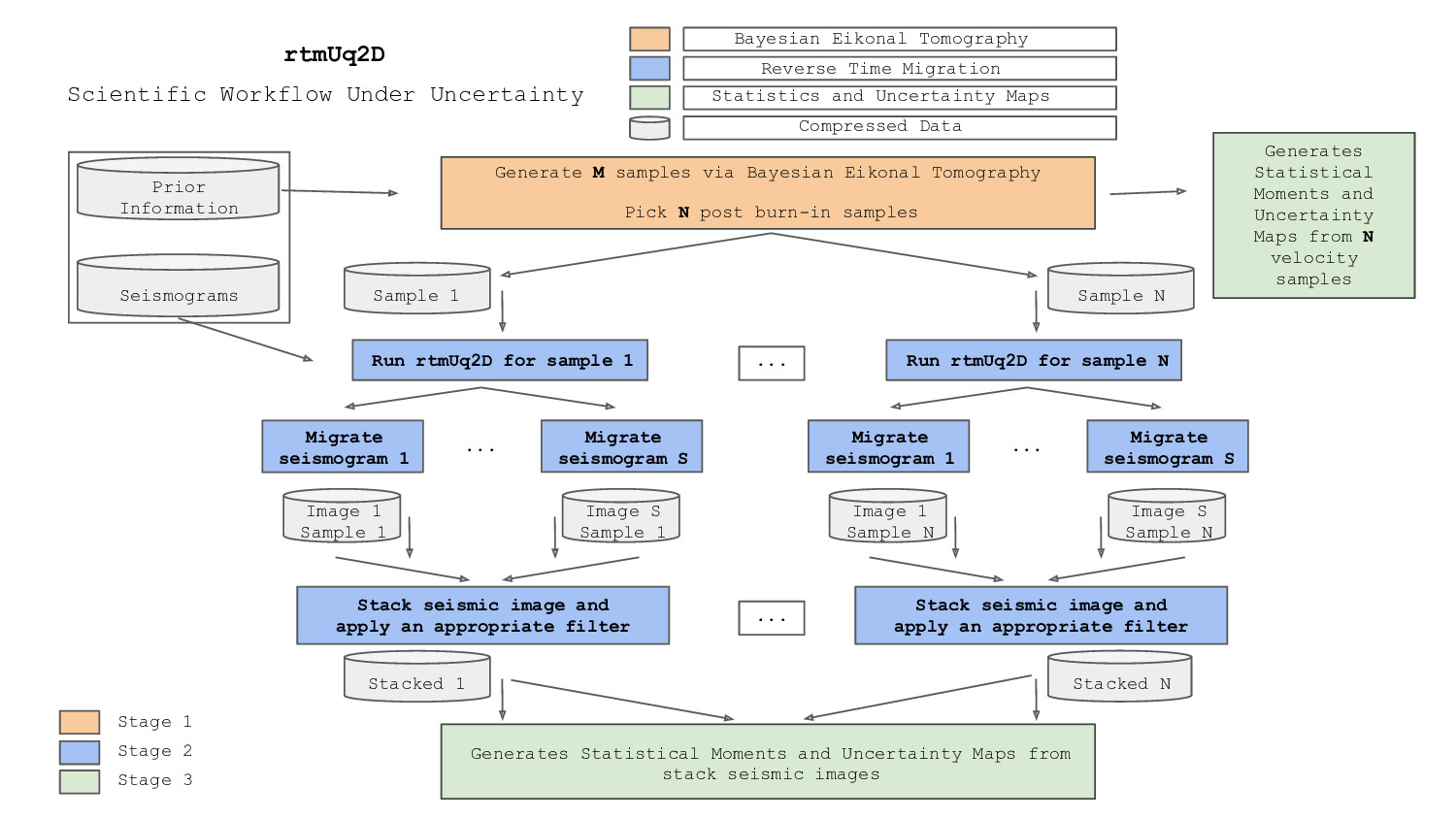}
\caption{Stage 1 provides a probabilistic distribution of velocity using BET. Stage 2 migrates the seismograms for all velocity parameters, generating, thus, a set of stacked seismic sections. Finally, Stage 3 extracts statistical information from velocity parameters and migrated images. Compression is used to transfer the data among the stages. Step 2 uses {\texttt{rtmUq2D} a hybrid parallel RTM code.}}
\label{FIG:WOKFLOWRTM}
\end{figure}

Figure \ref{FIG:WOKFLOWRTMIMAGES} shows the three workflow stages. BET produces an ensemble of velocity fields (stage 1), RTM generates the corresponding seismic images (stage 2), and stage 3 builds the uncertainty maps for the velocity and migrated images from the results of stages 1 and 2. To efficiently implement this workflow, we address three computational aspects: parallelism, data transfer, and data management. Figure \ref{FIG:WOKFLOWRTM} describes how the workflow produces the ensemble of velocities and migrated images for posterior uncertainty quantification, the hybrid parallelism in the RTM stage, and where data compression is used to reduce storage and network data transfer.

Stage 1 aims to build a probabilistic density function for the velocity encompassing feasible solutions that fit the calculated travel times to the data and provides enough information for the next two stages. To generate samples of the stochastic field, BET needs to run thousands of RJMCMC iterations. In the Voronoi's tessellation within RJMCMC the number of cells, their position and size vary. This strategy produces a number of Voronoi cells smaller than the velocity parameters, as a cell might contain many computational grid points. We use a KD-tree \citep{rawlinson2014seismic} to reduce the searching cost in the Voronoi's tessellation updating. Solving the Eikonal wave equation in each iteration dominates BET costs. We use here the open-source software FaATSO\footnote{\textbf{Fa}st Marching \textbf{A}coustic Emission \textbf{T}omography using \textbf{S}tandard \textbf{O}ptimization: https://github.com/nbrantut/FaATSO} to solve the Eikonal wave equation.

At the end of stage 1, we select a set of velocities for the migration based on the variance of successive batches of the post-burn-in samples. Thus, when the stochastic process converges, the variance among the samples stabilizes. In the second stage, RTM migrates the seismograms for the sub-set of velocity fields. These migrations produce the corresponding set of seismic images, where each one has a direct relation with one velocity sample coming from stage 1. To attenuate low-frequency artifacts in the brute stack images, we apply a Laplacian filter. The RTM kernel, that is, the second-order wave equation, is discretized with a second-order finite difference scheme in time and an eighth-order scheme in space. The reverse problem implementation explores the technique proposed by \cite{Givoli2014}. This technique guarantees the coercivity of the adjoint problem. Furthermore, the RTM code, {\texttt{rtmUq2D}}, takes advantage of the Single-Instruction-Multiple-Data (SIMD) model and memory alignment allocation to ensure vectorization. {\texttt{rtmUq2D}} also contains OpenMP directives to explore multiple cores parallelism. To complete our hybrid RTM solution, we apply MPI to manage the execution of multiple migrations and shots. Thus, we broke the total number of migrations into subgroups, and each subgroup is assigned to an MPI process, as exemplified in Figure \ref{FIG:WOKFLOWRTM}. When a subgroup finishes its workload, the next one starts immediately. The batch execution (an RTM for each subgroup) is repeated until exhausting the number of samples. Besides, each RTM verifies its own CFL condition to guarantee the stability of the discrete solution of the two-way wave equation for each sample.

Once stages 1 and 2 produce ensembles of velocity fields and seismic images, a condensed representation of this information can be calculated. In stage 3, velocity uncertainty maps are obtained as soon as we accept BET solutions. The same procedure is valid for generating migrated images. Hence, stage 3 generates uncertainty maps representing variability in velocity, and  reflectors amplitudes. These outcomes give to geoscientists extra information to guide the seismic interpretation.

However, to better support interpretation and improve confidence, it is essential to register relationships between data produced at each workflow stage. For example, the uncertainty maps of stage 3 are related to the velocity fields and statistics, which are also related to the seismic images. Tracking these data derivations among the stages at runtime, combined with the procedures' execution time, can support workflow parameters debugging and fine-tuning. Provenance data \citep{DavidsonFreire2008} represents the data derivation path of the data transformations during the stages. Provenance data capture and storing in a provenance database while the workflow executes can add an overhead to the workflow execution. We choose DfAnalyzer as a provenance asynchronous data capture software library because it can be used by high performance computing (HPC) workflows to provide for runtime data analysis with negligible overhead \citep{Silvaetal2018}. While monitoring the workflow execution, users submit queries to the travel time matrix during the migration or go directly from an input velocity model file to the corresponding files that have their statistics obtained after all samples were aggregated. DfAnalyzer has been successfully used for other HPC applications \citep{Camataetal2018, silva2016situ}.

Data transfer is also a concern when each stage in the workflow has to manage its information and send it to the next workflow stage. The amount of data transfer can increase drastically, given the number and size of the samples in the ensemble of velocities and migrated images. Data compression can reduce persistent storage and data transfer throughout the chain \citep{Lindstrometal2016}. The ZFP library is an open-source C/C++ library for compressing numerical arrays. To achieve high compression rates, ZFP uses lossy but optionally error-bounded compression. Although bit-for-bit lossless compression of floating-point data is not always possible, ZFP also offers an accurate near-lossless compression \citep{Lindstrometal2016}. We employ ZFP to compress the numerical arrays that store the velocity parameters, imaging conditions, and the seismograms. This use is linked to the workflow stages to reduce the pressure on the network data transfer among the workflow stages and to reduce I/O to persistent storage. However, lossy compression has to be managed carefully, and with safe error-bounds. We test both lossy and lossless compression, and we measure the error on the final results.

\section{Computational Experiments} \label{sec:experiments}
We designed some computational experiments to assess the overall workflow behavior and performance. They consist of verifying RTM strong and weak scalability, measuring the numerical errors on the final results after applying data compression, and discussing the workflow uncertainty quantification outcomes using the data analysis tools.  We choose two datasets: a two-layer constant velocity field for scalability analysis, and one based on the 2-D Marmousi2 benchmark \citep{WileyMarfurt2006} for applying data compression and running the full workflow.

\subsection{Scalability Analysis}
\label{sec:hybridparallelism}

We choose a two-layer constant velocity model that is 3.0 km depth and 9.2 km in the horizontal direction for the scalability tests. The velocities for the two layers are 2000 m/s, and 3000 m/s. We generate the reference data set for RTM based on the true velocity field. The reference data set is the wavefield recorded by near-surface hydrophones (one seismogram) with a cutoff frequency of 30 Hz. The input velocity field for RTM is a smooth version of the synthetic velocity model.

\begin{figure}[ht]
\centering\includegraphics[scale=0.3]{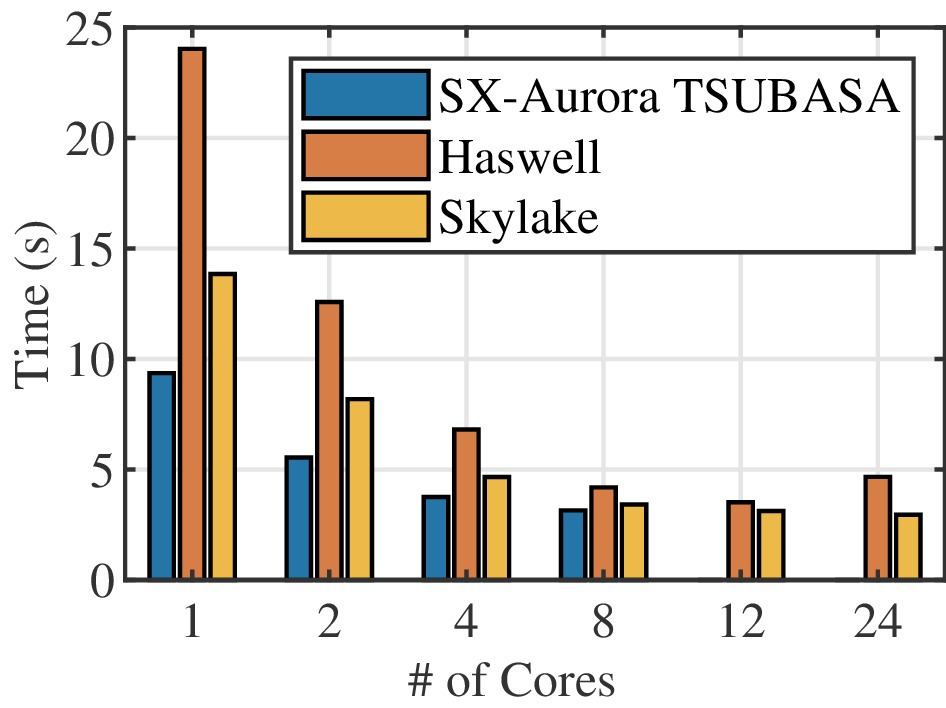}
\caption{RTM scalability analysis on different platforms. The tests consist of a migration for one seismogram. The vector processor available has 8 cores, while Haswell and Skylake have 24 cores each.}
\label{FIG:SCALABILITY}
\end{figure}

Figure \ref{FIG:SCALABILITY} shows the RTM strong scalability considering a migration of one shot for an increasing number of cores. We run the tests on different platforms: 1 vector processor (NEC SX-Aurora TSUBASA TypeB), and 2 multi-cores CPUs (Intel Xeon E5-2670v3 (Haswell) and Intel Xeon Platinum 8160 (Skylake) CPUs). The vector processor available has its minimum configuration, that is, 8 cores, while Haswell and Skylake have 24 cores each.

The code is portable across the different platforms requiring only the necessary compiler options for each system to enable aggressive optimization. Compiler flags are invoked for NEC vector instructions in SX-Aurora TSUBASA, Intel AVX2 vector instructions in Haswell, and Intel AVX512 instructions in Skylake. In all systems, the wave-equation kernels are 100\% vectorized. We can see that RTM on the vector processor runs faster than in Haswell and Skylake platforms, without a specific optimization effort for SX-Aurora TSUBASA.

Now we investigate the weak scalability. Due to our need to run the RTM $n$ times to build the statistics outcomes, the hybrid RTM has to run $n$ times in $n$ nodes. In this case, the CPU time for each node has to be approximately the same. Assuming the same parametrization of the strong scalability test, we run the RTM code for an increasing number of nodes and measure the CPU time for each one. We note that the RTM takes, on average, 3.281s with standard deviation 0.0578s to execute one RTM per node on the Haswell cluster. Thus, the CPU time for running one RTM in one node is approximately the same as running five RTM in five nodes. We observe the same pattern for the other platforms.

\subsection{Data Compression}\label{sec:compression}

We select the 2-D Marmousi2 velocity model benchmark \citep{WileyMarfurt2006} for the data compression experiments. The benchmark is 3.5 km depth and 17.0 km in the horizontal direction and has parameters models (p-wave velocity and density), and other data sets related to its acquisition. However, we generate the seismograms solving the two-way wave equation with true velocity. Near-surface hydrophones record the seismogram with a cutoff frequency of 45 Hz. Thus, we have 160 seismograms to be used in the migration stage. In the BET stage, the reference data set is 34 travel-time panels obtained from solving the Eikonal equation with true velocity.

After BET, we pick 5000 post-burn-in velocity models for migration. Thus, the RTM produced 5000 seismic images that corresponding with each velocity model from stage 1. Hence, we use 5000 velocity models, 5000 seismic migrations, and 160 seismograms to test the lossy and lossless rate compression. Table \ref{tbl:compressionVelMig} shows the results related to compression efficiency for the velocity models and seismic images. The compression efficiency for the seismograms are in Table \ref{tbl:compressionSeis}.

\begin{table}[ht]
\centering
\caption{Fixed accuracy lossy and lossless compression comparison. The error tolerances for lossy compression are $10^{-4}$, and $10^{-1}$. Each value is the average of the 1000 compressed matrices (velocity fields, and seismic images).} \label{tbl:compressionVelMig}
\begin{tabular}{l l l}
\hline
Compression & Velocity  (MB) & Image (MB) \\
\hline
None                & 6.10 & 6.10 \\
Lossless            & 4.78 &  5.34 \\
Lossy ($10^{-4}$)   & 4.77 &  3.65 \\
Lossy ($10^{-1}$)   & 2.98 & 1.83 \\
\hline
\end{tabular}
\end{table}

Each velocity model and the seismic image without compression has 6.1 MB. The values in the columns "Velocity"  and "Image"  represent, respectively, the average size of 1000 matrices for the velocity models and seismic images without compression (first line), and with lossless and lossy compression. We apply three compression types: the first one is lossless compression, the second one is lossy compression with an absolute error of $10^{-4}$, and the last one is lossy compression with an absolute error of $10^{-1}$. We can see that the lossy compression achieves the best compression rate, reaching about 51.1\% of compression for the velocity models (2.98 MB), and about 70.0\% of compression for the seismic images (1.83 MB). Note that the compression rates for each matrix are different since ZFP explores the structure of the data. Although less efficient, the lossless compression reaches about 21.6\% of compression for the velocity models (4.78 MB) and about 12.5\% of compression for the seismic images (5.34 MB). This efficiency rate for lossless compression is compatible with its purpose, which is to preserve the original information.

We also apply lossless and lossy compression to the 160 seismograms. Each seismogram has 76.20 MB. The error tolerance for lossy compression is to $10^{-4}$. Compression rates reach a minimum  of 13.38 MB and a maximum of 8.38 MB, as we can see in Table \ref{tbl:compressionSeis}. These values represent 82.4\%, and 89.0\% less than the original data, respectively.

\begin{table}[ht]
\centering
\caption{Seismograms fixed accuracy lossy and lossless compression comparison. The error tolerance for lossy compression is set to $10^{-4}$. The values represent the minimum and maximum compression among the 160 seismograms.}\label{tbl:compressionSeis}
\begin{tabular}{l l l}
\hline
Compression Type & Minimum  (MB) & Maximum (MB) \\
\hline
None & 76.20 & 76.20  \\
Lossless &  59.28 & 48.46  \\
Lossy & 13.38 & 8.38  \\
\hline
\end{tabular}
\end{table}

\begin{figure}[ht]
\centering \includegraphics[scale=.45]{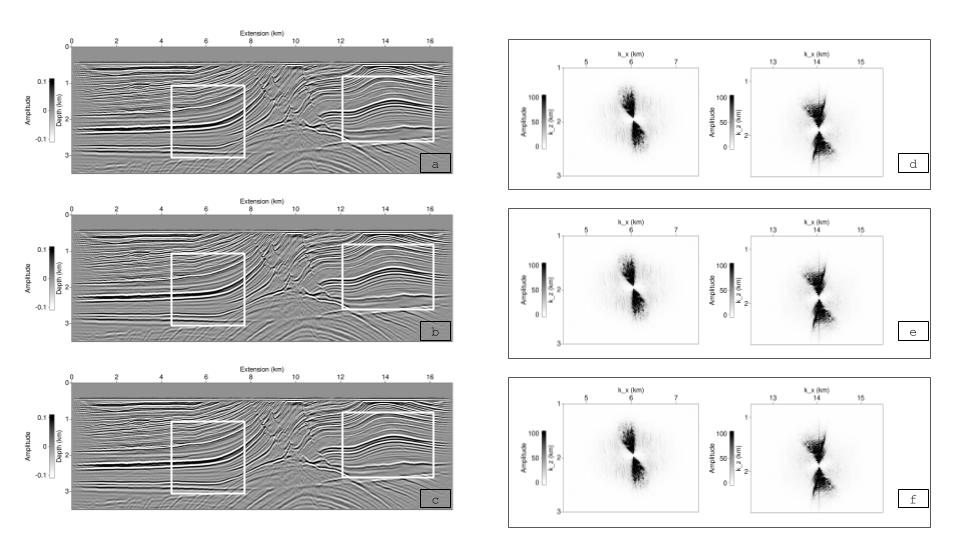}
\caption{RTM outcomes after migrating the (a) reference seismograms, seismograms with (b) lossless compression, and seismograms with (c) lossy compression. The velocity field inputs are not compressed for migration in Figure (a), losslessly compressed for migration in Figure (b), and lossy compressed for Figure (c). Figures (d), (e), and (f) represent the 2D Fourier spectra of the white box regions shown in Figures (a), (b), and (c), respectively.}
	\label{FIG:BZFPVELSEIS}
\end{figure}

The results from Table \ref{tbl:compressionVelMig}, and Table \ref{tbl:compressionSeis} suggest that we can use compressed data on the workflow and, thus, reduce the storage and network data transfer. However, we need to measure the error propagation due to compression on the final seismic images. Therefore, to measure the compression errors, we run RTM to migrate the seismograms without compression with an uncompressed velocity model as a parameter. Figure \ref{FIG:BZFPVELSEIS}(a) shows the migration of the uncompressed seismograms, and Figure \ref{FIG:BZFPVELSEIS}(d) shows the 2D Fourier spectra of the two white boxes shown in Figure \ref{FIG:BZFPVELSEIS}(a). The described outcomes are the reference once we do not use ZFP compression. The same procedure is applied using the lossless compression of the velocity model and seismograms as entries for RTM migration, and the lossy compression (tolerance error of $10^{-4}$) of the velocity model and seismograms as entries for RTM migration. These results can be seen in Figures \ref{FIG:BZFPVELSEIS}(a)(e), and Figures \ref{FIG:BZFPVELSEIS}(c)(f), respectively.

We calculate the normalized root mean square (NRMS) error  among the RTM results from Figure \ref{FIG:BZFPVELSEIS}(a), and the RTM results from Figures \ref{FIG:BZFPVELSEIS}(b), and \ref{FIG:BZFPVELSEIS}(c). The NRMS errors are $2.0 \times 10^{-6}\%$ and $4.3 \times 10^{-5}\%$, respectively. The NRMS errors among the spectra from the left white box regions in Figures \ref{FIG:BZFPVELSEIS}(a), \ref{FIG:BZFPVELSEIS}(b) and \ref{FIG:BZFPVELSEIS}(c) are $1.1 \times 10^{-6}\%$, and $1.97 \times 10^{-5}\%$, respectively. Moreover, the NRMS errors among the spectra from the right white box regions in Figures \ref{FIG:BZFPVELSEIS}(a), \ref{FIG:BZFPVELSEIS}(b) and \ref{FIG:BZFPVELSEIS}(c) are $1.3 \times 10^{-6}\%$, and $2.6 \times 10^{-5}\%$, respectively. The results from Figure \ref{FIG:BZFPVELSEIS} and the NRMS errors show that the chosen compression levels for the velocity models and seismograms lead to migrated images with errors of order $10^{-5}$\%. The compression strategy allows to reduce 85\% in storage, and, consequently, improves network data transfer without compromising accuracy.

\subsection{Provenance Data Queries}\label{sec:provenance}

Provenance data is captured to analyze data from the several workflow stages by DfAnalyzer and stored in a columnar database using the MonetDB  system\footnote{https://www.monetdb.org}. The workflow needs to have a definition of which data should be part of the provenance database. Then, DfAnalyzer calls are inserted in the workflow code to capture data provenance, as shown in Listing 1. We run a series of RTMs with and without DfAnalyzer calls.  In all tests, the overhead of provenance data capture and storage remains below 2.95\%, as shown in Table \ref{tbl:overheadProvenance}. Also, the overhead rate decreases as the number of nodes increases. During and after the workflow executions, queries can be submitted to the provenance database using filters or data aggregations over the workflow stages data derivation path, as shown in Figure \ref{FIG:wkfquery}.

\begin{table}[ht]
\centering
\caption{RTM overhead with DfAnalyzer calls. Each node executes two RTMs.}\label{tbl:overheadProvenance}
\begin{tabular}{l l l}
\hline
\# of Nodes & \# of RTMs & Overhead  (\%) \\
\hline
1    & 2 & 2.95  \\
2    & 4 & 2.63  \\
4    & 8 &1.26  \\
5    & 10 & 1.02  \\
\hline
\end{tabular}
\end{table}

\begin{figure}[ht]
\resizebox{\columnwidth}{!}{
    \begin{tabular}{|c|c|c|c|c|}
\hline
\textbf{shot} & \textbf{\begin{tabular}[c]{@{}c@{}}shot \\ Location\\ (x) m\end{tabular}} & \textbf{\begin{tabular}[c]{@{}c@{}}shot \\ Location\\ (z) m\end{tabular}} & \textbf{seismogram\_path} & \textbf{cross\_correlation\_path} \\ \hline
1 &  25.00 & 25 & ./.../seisNorm\_001.bin & ./.../crossCorre\_001X001.bin \\ \hline
2 & 131.25 & 25 & ./.../seisNorm\_002.bin & ./.../crossCorre\_002X001.bin \\ \hline
3 & 237.50 & 25 & ./.../seisNorm\_003.bin & ./.../crossCorre\_003X001.bin \\ \hline
4 & 243.75 & 25 & ./.../seisNorm\_004.bin & ./.../crossCorre\_004X001.bin \\ \hline
5 & 450.00 & 25 & ./.../seisNorm\_005.bin & ./.../crossCorre\_005X001.bin \\ \hline
6 & 556.25 & 25 & ./.../seisNorm\_006.bin & ./.../crossCorre\_006X001.bin \\ \hline
7 & 662.50 & 25 & ./.../seisNorm\_007.bin & ./.../crossCorre\_007X001.bin \\ \hline
8 & 768.75 & 25 & ./.../seisNorm\_008.bin & ./.../crossCorre\_008X001.bin \\ \hline
9 & 875.00 & 25 & ./.../seisNorm\_009.bin & ./.../crossCorre\_009X001.bin \\ \hline
10 & 981.25 & 25 & ./.../seisNorm\_010.bin & ./.../crossCorre\_010X001.bin \\ \hline
\end{tabular}}
	\caption{Provenance query result showing the shots with their ($x,z$) location and the path of their corresponding seismogram image file and the cross correlation file, filtered by a shot location where $z=25$ and $0< x <1000$.}
	\label{FIG:wkfquery}
\end{figure}

\subsection{Seismic Imaging Uncertainty Quantification}
\label{sec:uqmaps}

We discuss here the results of sequentially applying BET and RTM to the raw data provided by seismograms generating images with quantified uncertainty. The ensemble of uncertain velocity fields resulting from BET is assessed through primary statistics, such as velocity mean and standard deviation. In the sequence, we compute uncertainty maps. After the RTM migrates the seismograms for each velocity field, we analyze the impact of the inversion uncertainties in the seismic images. Note that in this stage, we choose to work on the compressed seismograms and velocity fields to reduce the storage and network transfer. As we show in Figure \ref{FIG:BZFPVELSEIS}, the lossy compression for both velocity and seismograms, with an error tolerance of $10^{-4}$, does not produce migrated seismic images with poor quality. Hence, we use them in the whole workflow. Besides, the Marmousi2 benchmark, as specified in Sub-section \ref{sec:compression}, is used to demonstrate the outcomes of the workflow. To keep track of the impact of each uncertainty source, we restrict the analysis in this particular example to the noise associated with monitoring the data, defined by $\sigma_{d}^{2}$ in equation \ref{eq.seistomo}. The other component that would induce uncertainty would be the scarcity of data promoted by acquiring signals only on the surface. That is circumvented by collecting the synthetic data in all grid points. We anticipate facing a lower degree of uncertainty in the final images due to such a choice.

\begin{lstlisting}[language=C++, caption=This workflow code shows DfAnalyzer calls inserted before RTM stage \texttt{runrtm.begin} to capture and register the input: velocityFile; wavelet and seismogram path with the corresponding output: cross-correlation file path after the RTM execution \texttt{runrtm.end} explicitly  relating input and output data in the provenance database.]
int task_id = 1;
Task task_runrtm = Task(dataflow.get_tag(), runrtm.get_tag(), task_id);

vector<string> irunrtm_values = {velocityFile, waveletFile, std::to_string(Nx), std::tostring(Nz)};
Dataset& ds_irunrtm = task_runrtm.add_dataset(irunrtm.get_tag());
ds_irunrtm.add_element_with_values(irunrtm_values);

//rtm per shot
for(shotID = 1; shotID <= numberOfShots; shotID++) {
    
    shotLocation [0] = firstLocationShot + (shotID - 1)*spacingShot;
    
    //forward modeling
    isotropicAcousticModeling (shotLocation, velocityModel, wavelet);
    
    sprintf(seismogramFile_02, seismogramFile, shotID);
    
    //begin DfAnalyzer input registration
    vector<string> iseismogram_values = {seismogramFile_02};
    Dataset& ds_iseismogram = task_runrtm.add_dataset(iseismogram.get_tag());
    ds_iseismogram.add_element_with_values(iseismogram_values);
    
    task_runrtm.begin();
    //end input registration
    
    reading_file(seismogramFile_02, seismogram, numberOfReceivers, numberOfTimeStep);
    
    //backward modeling
    adjointModeling(shotID, shotLocation, velocityModel, wavelet, seismogram);
  
    //begin DfAnalyzer output registration 
    if (shotID == numberOfShots)
        task_runrtm.end();
    else task_runrtm.save();
    //end output registration
}
\end{lstlisting}

\begin{figure}[ht]
	\centering
		\includegraphics[scale=.5]{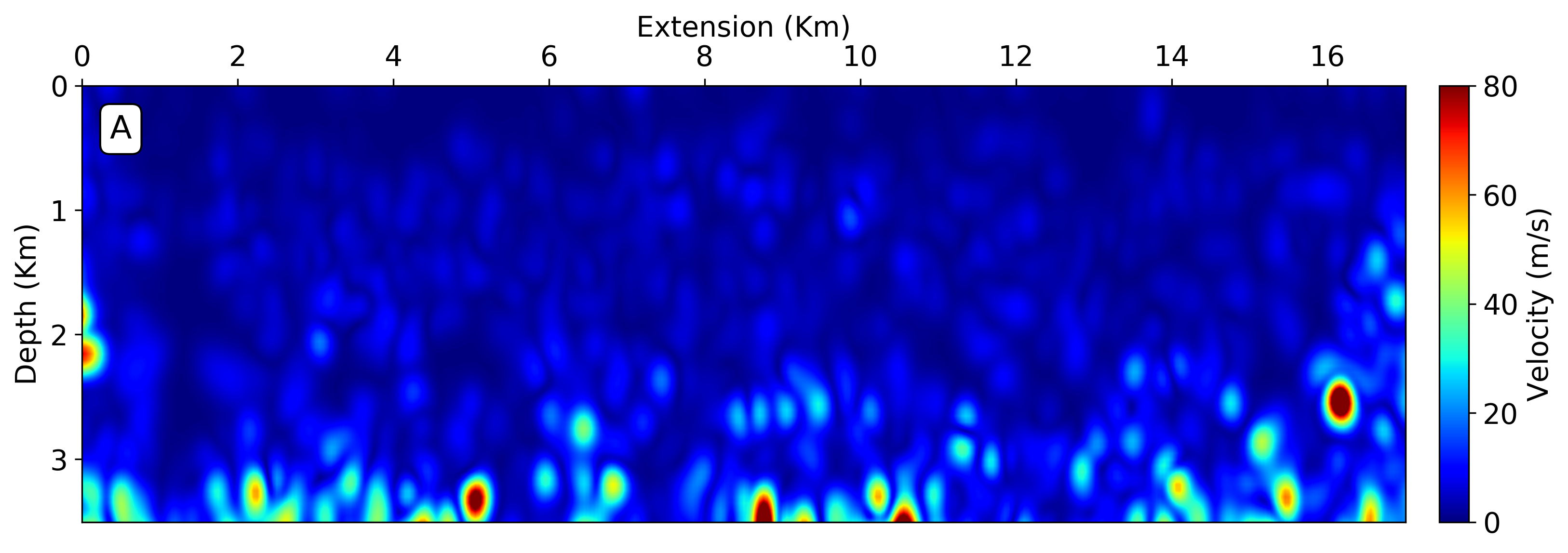}
		\includegraphics[scale=.5]{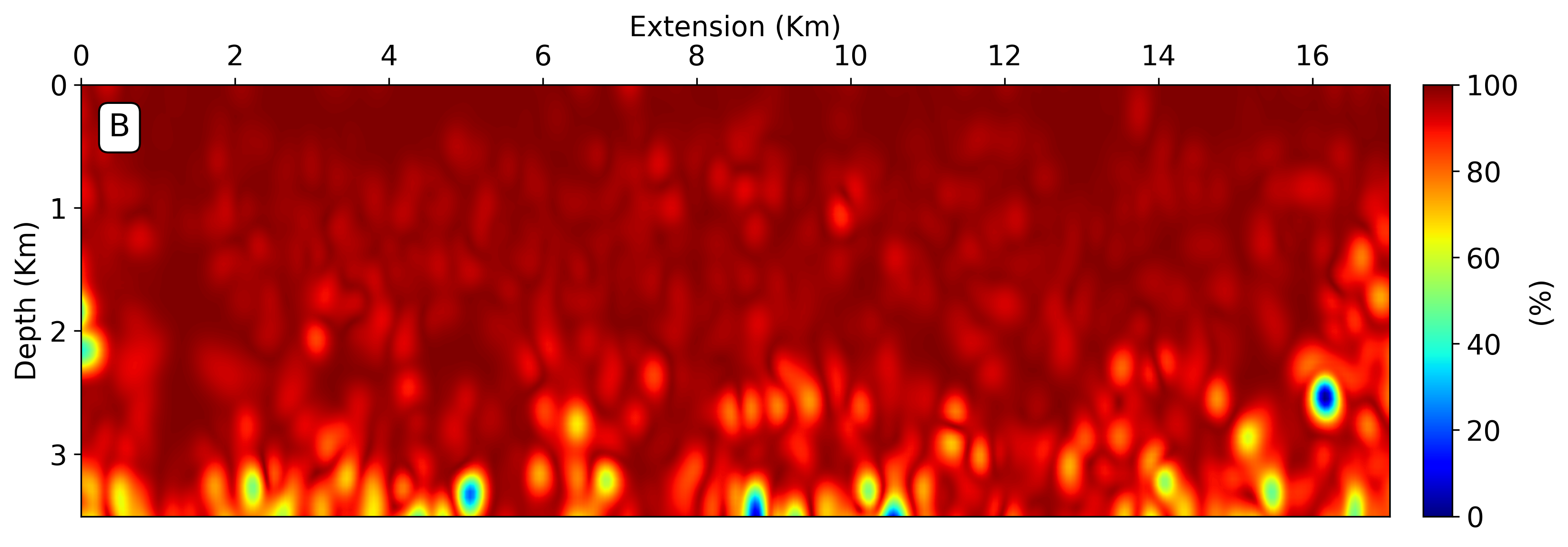}
		\includegraphics[scale=.5]{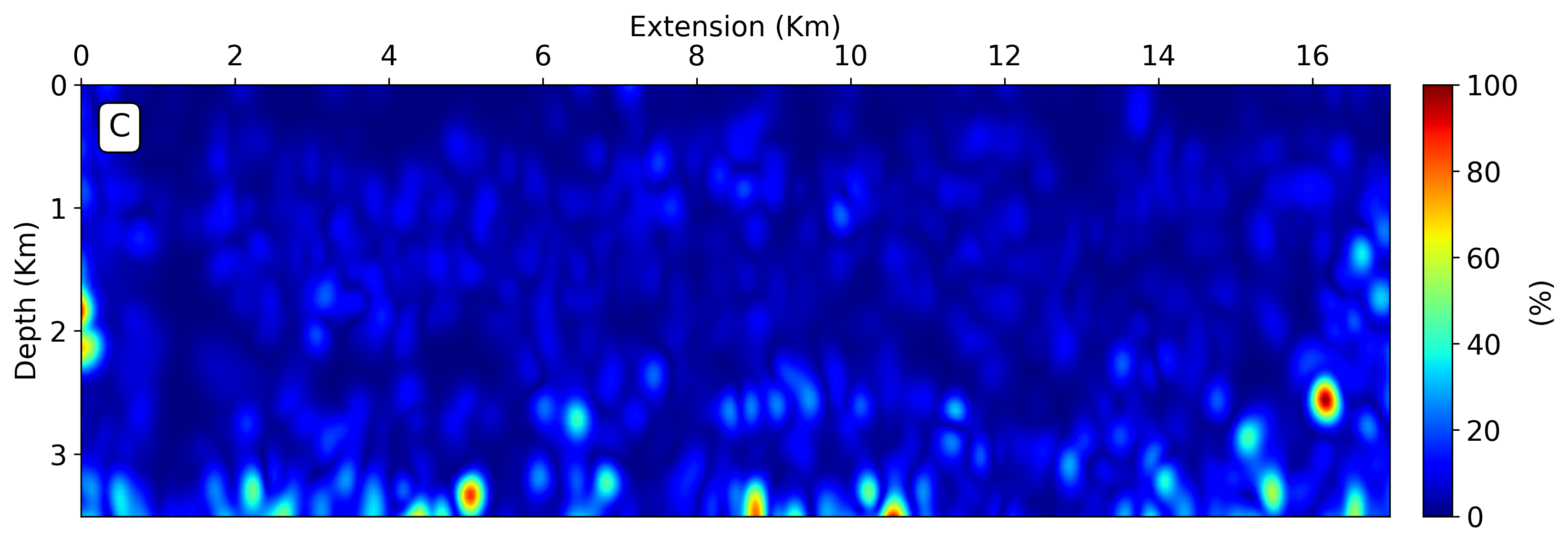}
	    \caption{Figures show the standard deviation (a), confidence index (b) and $r(\textbf{r})$ map (c) of the Marmousi velocity fields.}
	\label{FIG:UNCERTAINTYMAPS_VEL}
\end{figure}

Figure \ref{FIG:UNCERTAINTYMAPS_VEL} shows the standard deviation (a), confidence index (b), and standard deviation over the mean value (c) of the velocity fields. The standard deviation field expresses the degree of uncertainty in the velocity at each point of the domain. The minimum standard deviation values are around 1 m/s and localized in shallow regions. On the other hand, the maximum standard deviation values occur in deeper regions with values around 80 m/s. The confidence index  (Equation \ref{eq.confidence}) is the normalized standard deviation and represents the confidence degree in the velocities \citep{Lietal2015}. We depict in Figure \ref{FIG:UNCERTAINTYMAPS_VEL} (b) the confidence index within the image domain. We can see that such criterion assigns higher confidence ($> 0.8$) to shallow regions. Figure \ref{FIG:UNCERTAINTYMAPS_VEL} (c) shows $r(\textbf{r})$ map. Again we observe the same trends.

\begin{figure}[ht]
	\centering
		\includegraphics[scale=.5]{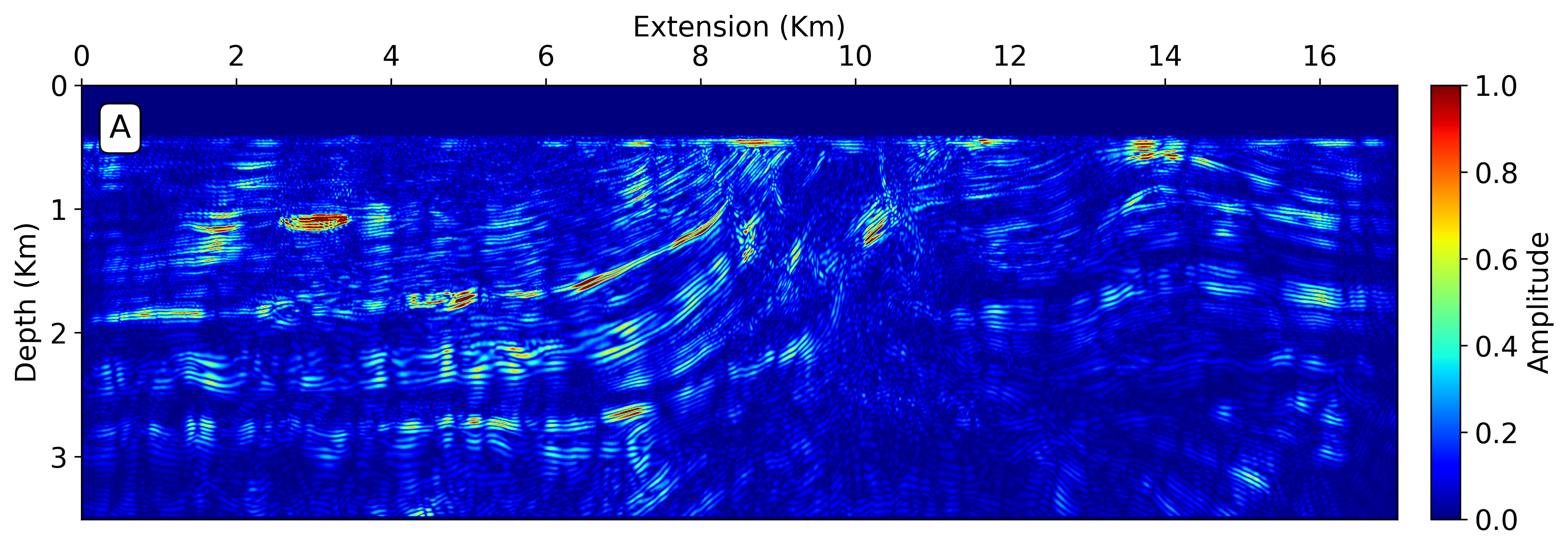}
		\includegraphics[scale=.5]{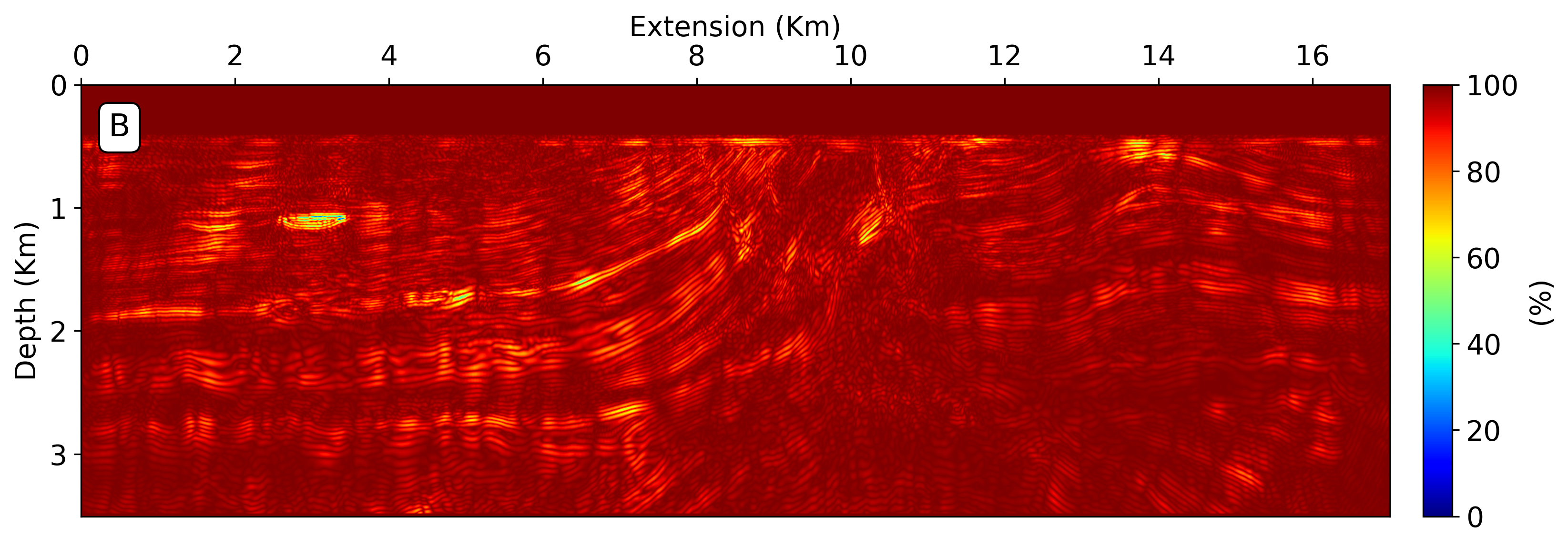}
		\includegraphics[scale=.5]{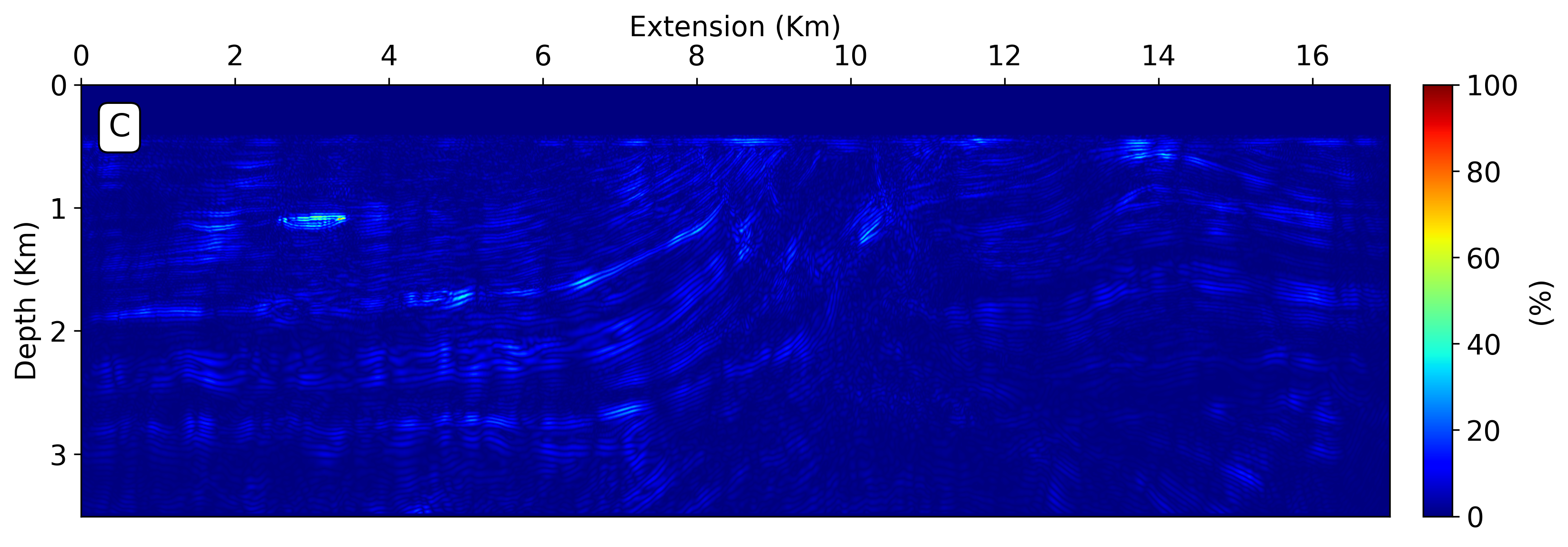}
	    \caption{Figures show the standard deviation (a), confidence index (b) and $r(\textbf{r})$ map (c) of the migrated seismic images. RTM migrates 160 seismograms for each velocity field.}
	\label{FIG:UNCERTAINTYMAPS_MIG}
\end{figure}

We plot in Figure \ref{FIG:UNCERTAINTYMAPS_MIG} (a) the standard deviation for a representative image computed as an average amongst those of the ensemble. Besides, we plot the confidence index (b) and $r(\textbf{r})$ map (c) of the seismic images obtained from RTM wrapped on the MCMC. In this case, the observed variations are related to the amplitude of the reflections instead of p-wave velocity. That is the variability of the amplitude intensity. This one is maximum in the interface between two rocks with different physical properties. However, if an estimated p-wave velocity differs from the exact p-wave velocity of the medium, the amplitude intensity is misplaced. Hence, the high standard deviation, low confidence index, and low standard deviation over the mean values are highlighted in the interfaces and predominant in the middle region.

\section{Conclusions}
\label{sec:conclusions}

We proposed a new scientific workflow that supports uncertainty quantification in seismic imaging. Our stochastic perspective to handle uncertainties is composed of three sequential stages: inversion, migration, and interpretation. A Bayesian tomography (Stage 1) provides the input data for the RTM wrapped on the MCMC (Stage 2). The workflow deploys means to end-users to explore uncertainty in the final images through maps computed in Stage 3.

The workflow execution requires significant computing power and generates a huge amount of data. Thus, we use optimized hybrid computing solutions that explore vectorization, data compression, and advanced data management tools. These tools allow us to obtain a scalable solution for the Marmousi Velocity Model benchmark with quantified uncertainty. ZFP compression has been used to transfer information among the stages and to store the final images. High levels of compression have been observed for the ensembles of velocities and migrated images. The compression strategy improves the network data transfer and reduces the amount of data up to 85\%, assuming a tolerance error of $10^{-4}$ without affecting the quality of the final result. The provenance database keeps a record of the workflow execution stages enriching the seismic data analysis quality. We observed that capturing and storing provenance data during the workflow execution is inexpensive.

\section*{Acknowledgements}
This study was financed in part by CAPES, Brasil Finance Code 001. This work is also partially supported by FAPERJ, CNPq, and Petrobras. Computer time on Stampede is provided by TACC, the University of Texas at Austin. The US NSF supports Stampede under award ACI-1134872. Computer time on Lobo Carneiro machine at COPPE/UFRJ is also acknowledged. We are also indebted to NEC Corporation.

\section*{Computer Code Availability} 

The codes can be downloaded at the repository \url{https://github.com/Uncertainty-Quantification-Project}. The repository provides three opensource codes (\texttt{bayes-tomo}, \texttt{rtmUq2D}, and \texttt{rtmUq2D-Prov}). The user guide describes the software and hardware requirements, and directs to complementary opensource software used in the paper.






\bibliographystyle{elsarticle-num-names}
\bibliography{elsarticle-template-1-num.bib}







\end{document}